\documentclass{imamat}

\usepackage{amsmath}
\usepackage{amssymb}
\usepackage{graphicx}
\usepackage[letterpaper, margin=1in]{geometry}
\usepackage{subfig}
\usepackage{mathrsfs}
\usepackage{xcolor}
\numberwithin{equation}{section}

\begin{document}

\title{Gradient Induced Droplet Motion Over Soft Solids}

\shorttitle{Gradient Induced Droplet Motion Over Soft Solids} 
\shortauthorlist{Bardall, Chen, Daniels, Shearer} 

\author{
\name{Aaron Bardall$^*$}
\address{North Carolina State University Dept. of Mathematics\email{$^*$Corresponding author: arbardal@ncsu.edu}}
\name{Shih-Yuan Chen}
\address{North Carolina State University Dept. of Physics}
\name{Karen E. Daniels}
\address{North Carolina State University Dept. of Physics}
\and
\name{Michael Shearer}
\address{North Carolina State University Dept. of Mathematics}}

\maketitle

\begin{abstract}
{
Fluid droplets can be induced to move over rigid or flexible surfaces under external or body forces. We describe the effect of variations in material properties of a flexible substrate as a mechanism for motion.  
In this paper, we consider a droplet placed on a substrate with either a stiffness or surface energy gradient, and consider its potential for motion via coupling to elastic deformations of the substrate.
In order to clarify the role of contact angles and to obtain a tractable model, we consider a two-dimensional droplet. The gradients in substrate material properties give rise to asymmetric solid deformation and to unequal contact angles, thereby producing a force on the droplet. We then use a dynamic viscoelastic model to predict the resulting dynamics of droplets. Numerical results quantifying the effect of the gradients establish that it is more feasible to induce droplet motion with a gradient in surface energy. The results show that the magnitude of elastic modulus gradient needed to induce droplet motion exceeds experimentally feasible limits in the production of soft solids and is therefore unlikely as a passive mechanism for cell motion. In both cases, of surface energy or elastic modulus, the threshold to initiate motion is achieved at lower mean values of the material properties.
}
{Droplet motion, soft solids.}
\end{abstract}

\section{Introduction}
The deformation of a soft substrate induced by the presence of a resting fluid droplet raises issues of force balance at the contact line, 
  and how to determine  the shape of the substrate free surface. These issues have been addressed with experiments, modeling and theory \citep{r40,r24,r52,r17,r45,r15,r28}. A more challenging problem, addressed in this paper, is to characterize conditions under which the droplet can be set in motion and to determine the subsequent dynamics, specifically the droplet speed. 
In recent experiments in \citet{r5}, migration of fluid droplets is observed to be driven by a gradient in substrate thickness, a physical surrogate for the bulk elastic modulus.  It was found that droplets migrate to thicker, less hard regions of the substrate.

In {\em durotaxis,}  the motion of cells is induced by a rigidity gradient in the surrounding soft material 
\citep{r26}
In this process, the cells are thought to actively sense and relay changes in stiffness and respond by contracting and altering cellular shape to migrate to regions of higher stiffness. Interestingly, this direction of migration is opposite to the observations of droplet motion over substrates of varying thickness \citep{r5}, in which substrate thickness is considered to be a surrogate for bulk elastic modulus.  In the experiments reported in \citet{r5},  inorganic fluid droplets are observed to migrate towards thicker, less hard regions of the substrate.   On the other hand, recent computational results have suggested that fluid droplet migration across a true rigidity gradient may be biased toward stiffer regions of the substrate \citep{r4}, and may depend on the balance of interfacial energies at the fluid-solid-vapor contact line.  
Considering this mixture of evidence concerning durotaxis, analysis of modes of migration of fluid droplets will help resolve the issue, and potentially provide insight into the driving mechanisms for living cells.

  Droplet motion on {\em rigid} surfaces can be induced by periodically patterned surface energies \citep{r41,r37,r22}, thermal gradients \citep{r46}, and magnetic fields \citep{r42}.   In general, droplet motion (on rigid or  soft substrates), relies on contact angle asymmetry, resulting in a force imbalance that drives motion.  However, 
  droplet motion across  {\em soft} substrates also depends on elastic deformation, and the resulting elastic energy,   as well as bulk energy dissipation, within the substrate.  The analysis of these quantities is central to determining conditions under which a fluid droplet will passively migrate across a soft surface.  
 
 Throughout the paper, we consider a two-dimensional droplet. This simplification, though unphysical, allows us to analyze the contact angle asymmetry resulting from   nonuniformity in substrate properties. 
In \S\ref{asymmetric} we formulate equations for the substrate deformation due to the droplet, as well as due to gradients in substrate properties such as   elastic modulus and surface energy. Boundary conditions are formulated that help determine the free surface of the substrate and the structure of the contact line. The equations are analyzed using Fourier transforms. We show the dependence of contact angle asymmetry on the degree of substrate nonuniformity as a precursor to analyzing droplet motion in \S\ref{dynamics}.  In that section, we formulate a model that predicts droplet velocity, based on the rate of energy dissipation. Finally, in \S\ref{Discussion} we  summarize and interpret the results. In particular we deduce that for practical purposes, a gradient in surface energy is more likely to induce droplet motion than a gradient in elastic modulus.

\section{Substrate Gradients and Asymmetric Deformation}\label{asymmetric}

When a fluid droplet rests on a horizontal {\em rigid} substrate, surface energies $\gamma$ at the phase interfaces govern the equilibrium contact angle $\theta_Y$ of the droplet through Young's equation:
\begin{equation}\label{YoungsEq}
\gamma_{sg} - \gamma_{ls} = \gamma_{lg}\cos\theta_Y.
\end{equation}
 Here, subscripts on surface energy terms $\gamma$ refer to the associated interface,  e.g., $\gamma_{sg}$ is the surface energy of the solid-gas interface.  Young's equation results from minimization of the total surface energy  at the phase interfaces; it represents horizontal force balance at the contact line.  The remaining vertical force caused by the liquid-gas interface of the droplet is assumed to be resolved by the property of an ideal rigid solid substrate in which strains approach zero as the elastic modulus tends to infinity.  When the difference in solid surface tensions $\gamma_{sg} - \gamma_{ls}$ is positive, the droplet-substrate system is termed {\em hydrophilic,} with an equilibrium Young's angle $\theta_Y < 90^\circ$, whereas if the difference is negative, the droplet-substrate system is {\em hydrophobic,} with Young's angle $\theta_Y > 90^\circ$.
 
In contrast, for a fluid droplet resting on a  {\em soft} solids substrate, non-zero deformations occur in the substrate, influenced by the capillary forces of the droplet that 
 introduce elastic energy into the system.  This elastic energy competes with surface energy so that determination of the equilibrium contact angle via minimization of the total system energy becomes more complex.

 \begin{figure}
\centerline{\includegraphics[scale=2]{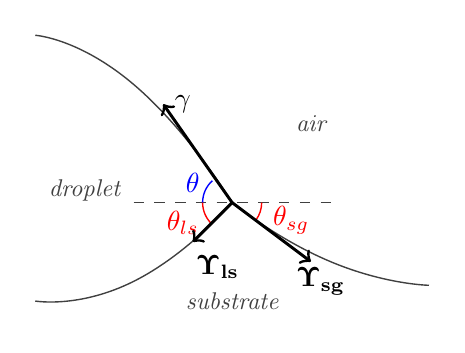}}
\caption[Wetting ridge formed by a resting fluid droplet]{Wetting ridge formed by a resting fluid droplet.  Solid surface stress $\Upsilon$ balances surface tension of the contact line $\gamma$.  Apparent contact angle $\theta$ illustrated in blue as the angle the fluid surface forms with the far field horizontal (dashed line), while solid contact angles $\theta_{ls}$ and $\theta_{sg}$ define respectively the angle of the liquid-solid and solid-gas interfaces form with the far field horizontal.}
\label{fig:tikz_Neumann}
\end{figure}

The vertical component of force from the liquid-gas interface pulls up on the substrate creating a wetting ridge, as illustrated in Fig.~\ref{fig:tikz_Neumann}.   The solid interfaces on either side of the triple point angle downward, opposing the upward pull of the droplet edge. This  creates a total force balance at the contact line quantified by  Neumann's triangle \citep{r15,r29}:
\begin{equation}\label{Neumann}
\vec{\Upsilon}_{sg} + \vec{\Upsilon}_{ls} + \vec{\gamma} = \vec{0}.
\end{equation}
In  this generalization of Young's equation, solid surface stress vectors $\vec{\Upsilon}$ balance   the surface tension vector $\vec{\gamma}$ of the droplet edge.  The wetting ridge  represents a deformation  magnitude  of approximately the   elastocapillary length   $L_e = \gamma/E$, where $E$ is the elastic modulus of the solid substrate.  With typical liquid surface energies $\gamma \approx 6 \times 10^{-2}$ N/m, we find that micro-scale deformations occur at elastic moduli $E$ of order $1-10~\text{kPa}$.  Note that in \citet{r54} it is shown that, though deformation is negligible in stiff substrates with elastocapillary length $L_e$ much less than one micrometer ($L_e \ll 1~\mu$m), if   $L_e$ is larger than the atomic length scale, then Neumann's triangle is still formed at the elastocapillary length scale despite visibly negligible deformation. %
 
For droplets of size comparable to the elastocapillary length $L_e$, substrate deformations are significant enough to alter the apparent contact angle $\theta$ of the droplet (shown in Fig.~\ref{fig:tikz_Neumann}) several degrees from Young's angle $\theta_Y$ in \eqref{YoungsEq} \citep{r15}.  While deviation from Young's angle causes total surface energy to increase, deviations that reduce the angle with which the liquid-gas interface meets the solid surface reduces the upward pull of the liquid edge resulting in shallower deformations and strains and thus lower elastic energy.  The competition between these two trends generally results in hydrophobic droplet-substrate systems having equilibrium angles slightly greater than  predicted by \eqref{YoungsEq}, and similarly results in hydrophilic systems having slightly smaller equilibrium angles  than   predicted by \eqref{YoungsEq} \citep{r4,r15}.  
 
When a contact line at the fluid-solid-vapor interface advances or recedes, there are associated advancing and receding contact angles $\theta_a$ and $\theta_b$ respectively such that $\theta_b \leq \theta \leq \theta_a$, where $\theta$ is the static contact angle of the droplet.  These dynamic contact lines exhibit hysteretic behavior. Specifically,  when  fluid  is added to the droplet, the contact angle will increase to the advancing angle $\theta_a$, and  the contact line will advance, eventually settling  when the  equilibrium contact angle is reached.  If the additional fluid is then removed, the contact angle  decreases down to the receding contact angle $\theta_b$ before the contact line recedes to achieve the droplet equilibrium shape \citep{r10}.  These angles can  be found experimentally using the tilted plate method, in which   a droplet is rested on a horizontal plate which is then  tilted  until the droplet begins to migrate.  The contact angles formed at the front and back of the droplet are   the advancing and receding contact angles respectively \citep{r10}.
 
By introducing a gradient in a substrate property such as elastic modulus or surface energy, the apparent contact angle of a resting droplet becomes spatially dependent, resulting in contact angle asymmetry.  With enough bias in one direction, the force imbalance generated by the asymmetry can overcome pinning forces at the contact line and droplet motion is induced.    Experimental results of \citet{r5}  determine that  a contact angle difference of $\Delta \theta = \theta_a - \theta_b \approx 1.8^{\circ}$ is sufficient for water droplet motion over a silicone gel substrate.  These results were independent of the droplet size, which is a factor in the equilibrium contact angle of the droplet on a soft surface.  This suggests that motion is governed by the relative difference in contact angles as opposed to the advancing and receding angles themselves.  
  
We consider a droplet on a soft substrate with either a gradient in elastic shear modulus $G(x)$ or a gradient in solid surface energy $\gamma_s(x)$.  We  focus on incompressible substrates, for which the Poisson's ratio is $\nu = 1/2$.  In addition, for simplicity we assume that the strain dependence of the surface stress is negligible compared to the mean surface stress.  This allows us to set the surface stress $\Upsilon(x)$ equal to $\gamma_s(x)$.  Furthermore we restrict the analysis to that where the solid surface energy $\gamma_s(x)$ is assumed to be continuous, representative of droplets with Young's angle $\theta_Y = 90^\circ$.  For the case of applying a gradient in shear modulus $G(x)$, we will take the solid surface energy $\gamma_s$ to be constant and in the case of the surface energy gradient we will take shear modulus $G$ to be constant.
 
By allowing spatially dependent substrate properties such as modulus $G$ and surface energy $\gamma_s$, the static deformation caused by a resting fluid droplet becomes asymmetric.  Altering these properties also affects the apparent contact angle of the droplet \citep{r11,r15,r5}. Thus, allowing stiffness or surface energy to become spatially dependent in general allows the two dimensional droplet to have an asymmetric profile and a nonzero difference in contact angle from left to right.  This contact angle difference determines the onset of droplet motion \citep{r10,r5} and we will use the magnitude of contact angle difference from \citet{r5} of 1.8$^\circ$ as a benchmark for inducing droplet motion.  We acknowledge that this threshold may be a function of the liquid-solid system but, as indicated by the results of \citet{r5}, it is not a function of the droplet size.
 
To formulate equations for deformation of the substrate, we define a reference configuration 
$\Omega=\{(x,z): -\infty<x<\infty, 0<z<h\},$ representing the substrate with no deformation. Deformation is then given by a mapping $(u,w),$ so that the deformed substrate in physical space is $\{(x+u(x,z),z+w(x,z)): (x,z)\in \Omega\}, $
and the substrate free surface is $\{(x+u(x,h),h+w(x,h)), |x|<\infty\}.$ We assume the droplet covers the portion $\{(x,h), -R<x<R\}$ of the free surface, in the reference configuration.
 
Since deformations within the substrate are small, we assume linear elasticity, for which the stress-strain relation is written
 \begin{equation}\label{inc_stress_tensor}
\tau_{ij}(x,z) = 2G\varepsilon_{ij}(x,z) - p(x,z)\delta_{ij}, \quad i,j=1,2,
\end{equation}
where $G=G(x)$ is the shear modulus of the substrate (independent of $z$) and $p$ represents the isotropic stress in the solid.  Along with this stress tensor, for a finite value of $p$ we enforce the incompressibility condition that the strain tensor $\varepsilon$ (the symmetric part of $\nabla(u,w)$) is trace-free:
\begin{equation}\label{trace_strain}
  \varepsilon_{xx} + \varepsilon_{zz} = \partial_xu + \partial_zw = 0.
\end{equation}
Note that subscripts $i,j =1,2$ are used interchangeably with $x,z$ in notation for stress and strain tensors. 
Stress boundary conditions due to the effect of the droplet on the free surface in general allow two different contact angles $\theta_l$ and $\theta_r$ (on the left and right, respectively).  Since the droplet contact angles may differ from Young's angle $\theta_Y$ due to elastic deformation at the wetting ridge,  tangential contact line forces $f_l$ and $f_r$ are given by
\begin{equation}\label{flr}
f_{l,r} = \gamma(\cos\theta_{l,r} - \cos\theta_Y).
\end{equation}
Then stress boundary conditions at the substrate free surface are:
\begin{subequations}\label{stressBCstatic}
\begin{alignat}{2}
 \tau_{xz}|_{z=h} &= f_l\delta(x+R) - f_r\delta(x - R) + k^2\Upsilon\partial_{xx}u|_{z=h}\\
\tau_{zz}|_{z=h} &= \gamma\sin\theta_l\,\delta(x+R) + \gamma\sin\theta_r\,\delta(x-R) - \Pi\,H(R-|x|) + \Upsilon\partial_{xx}w|_{z=h}
 \end{alignat}
\end{subequations}
Here, $\gamma$ is the fluid surface stress, $\Upsilon=\gamma_s$ is the solid surface stress, $\Pi$ is the fluid pressure underneath the droplet, and $\delta$ and $H$ are respectively the Dirac-delta and Heaviside distributions.
The parameter $k^2$ relates to the curvature of the free surface, as outlined in \citet{r24}.
 
We employ a Fourier transform approach, generalized to allow for asymmetric contact angles and deformation, to solve the equilibrium equation
\begin{equation}\label{modeleq}
\nabla \cdot \bar{\tau} = \vec{0},
\end{equation}
 with fixed boundary conditions at the base of the substrate ($u(x,0) = w(x,0) = 0$),
 and the stress boundary conditions \eqref{stressBCstatic} at the free surface $z=h.$  
 
Next we compute the elastic energy in the substrate for the calculated deformation field. The  total energy functional,   the sum of the elastic and surface energies, depends on   static contact angles $\theta_{l,r}$ that appear in the boundary conditions. Consequently, by minimizing the total energy we determine the the static contact angles $\theta_{l,r}$ for given surface stress or substrate stiffness distributions. 

\subsection{Surface Energy Gradient}\label{s5p2}
In this subsection, we solve the model equations for the case of a gradient in solid surface energy $\gamma_s$.  Under our assumptions we have that $\Upsilon(x) = \gamma_s(x)$, and we prescribe a small perturbation to the average surface energy $\bar{\gamma}_s$:
\begin{equation}\label{Yvar}
\gamma_s(x) = \bar{\gamma}_s + a\tilde{\gamma}_s(x) \qquad \tilde{\gamma}_s(x) = \frac{2}{\pi}\arctan(x/L)
\end{equation}
where  $a$ is a small parameter so that the total variation in surface energy is much less than the average surface energy ($|a| \ll \bar{\gamma}_s$), and $L$ controls the length over which the surface energy gradient is present.  In order to solve the model equations $\partial_j\tau_{ij} = 0$, we utilize an expansion in small parameter $a$ and apply a Fourier transform in the horizontal direction on the scale separated boundary value problems.  We expand with respect to small parameter $a: p=p_0+ap_1+ \mathcal{O}(a^2),$ etc.  Applying scale separation to the force balance equations, we  have, at $\mathcal{O}(1):$
 
\begin{subequations}\label{BVP0_Y}
\begin{alignat}{4}
\nabla p_0 \ &= \ G\Delta\langle u_0,w_0\rangle, \label{pde0}\\
 \big[G(\partial_zu_0 + \partial_xw_0) - k^2\bar{\gamma}_s\partial_{xx}u_0\big]_{z=h} &= \ f_l\delta(x+R) - f_r\delta(x-R)\label{BVP0_bc1Y} \\
\big[2G\partial_zw_0 - p_0 - \bar{\gamma}_s\partial_{xx}w_0\big]_{z=h} &=\ \gamma\big(\sin\theta_l\delta(x+R) + \sin\theta_r\delta(x-R)\big) - \Pi H(R-|x|)\label{BVP0_bc2Y},
\end{alignat}
\end{subequations}
and at $\mathcal{O}(a):$
 \begin{subequations}\label{BVP1_Y}
\begin{alignat}{4}
\nabla p_1 \ =& \ G\Delta\langle u_1,w_1\rangle, \label{pde1}\\
\big[G(\partial_zu_1 + \partial_xw_1) - k^2\bar{\gamma}_s\partial_{xx}u_1\big]_{z=h} =& \ \big[k^2\tilde{\gamma}_s\partial_{xx}u_0\big]_{z=h}\label{BVP1_bc1Y}\hspace{2.03in}\\
\big[2G\partial_zw_1 - p_1 - \bar{\gamma}_s\partial_{xx}w_1\big]_{z=h} =& \ \big[\tilde{\gamma}_s\partial_{xx}w_0\big]_{z=h}.\label{BVP1_bc2Y}
\end{alignat}
\end{subequations}
Note that  the scale separated equations \eqref{pde0}, \eqref{pde1},    together with \eqref{trace_strain}, have the same  homogeneous PDE structure.  

We define the Fourier transform pair to be
\begin{align*}
\mathcal{F}\big[f(x)\big] =& \int_{-\infty}^\infty f(x) e^{-isx} dx = \hat{f}(s), \quad
\mathcal{F}^{-1}\big[\hat{f}(s)\big] = \frac{1}{2\pi} \int_{-\infty}^\infty \hat{f}(s) e^{isx} ds = f(x).
\end{align*}
 Transforming the system \eqref{trace_strain}, \eqref{pde0}, \eqref{pde1}  and eliminating $p_i, i=0,1$,  we obtain ordinary differential equations governing the transformed displacements:
$$(\partial_{zz}-s^2)^2\hat{w}_j = 0, \qquad \hat{u}_j = is^{-1}\hat{w}_j, \qquad (j = 0,1).$$
 The general solution satisfying the boundary conditions at $z=0,$  takes the form
\begin{equation}\label{uwjY}
\hat{u}_j(s,z) = iC_j(s)\psi'(sz) + iD_j(s)\psi''(sz),\quad
\hat{w}_j(s,z) = C_j(s)\psi(sz) + D_j(s)\psi'(sz)
\end{equation}
where
\begin{equation}\label{psidef}
\psi(\xi) = -\sinh(\xi) + \xi\cosh(\xi), \quad \psi^{(k)}(sz) = \frac{d^k}{d\xi^k}\psi(\xi)|_{\xi = sz} = s^{-k}(\partial_z)^k\psi(sz), \quad k = 0,1,2,\cdots.
\end{equation}
The Fourier coefficients $C_j(s)$, $D_j(s)$ ($j=0,1$) are obtained by transforming the shear and normal boundary conditions   \eqref{stressBCstatic}, leading to the 
linear system of equations:
\begin{subequations}\label{CDj_asym}
\begin{alignat}{2}
C_j(s)\beta(s) + D_j(s)\beta^{\diamond}(s) &= M_j(s), \quad 
C_j(s)\mu(s) + D_j(s)\mu^{\diamond}(s) &= N_j(s),
\end{alignat}
\end{subequations}
where the right hand sides of the linear system are defined as
\begin{subequations}\label{MNY}
\begin{alignat}{4}
M_0(s) &= (f_r + f_l)\sin sR + i(f_r - f_l)\cos sR, \label{M0}\\
N_0(s) &= \gamma(\sin\theta_l+\sin\theta_r)\cos sR + i\gamma(\sin\theta_l-\sin\theta_r)\sin sR - 2\Pi\frac{\sin sR}{s}, \label{N0} \\
 M_1(s) &= \frac{ik^2}{2\pi}\Bigg(\frac{\mathcal{F}\big[\tilde{\gamma}_s\hspace{.001in}'
\big]}{s}\Bigg)*\big(s\partial_z\hat{w}_0\big)_{z=h}, \quad
N_1(s) = \frac{i}{2\pi}\Bigg(\frac{\mathcal{F}\big[\tilde{\gamma}_s\hspace{.001in}'\big]}{s}\Bigg)*\big(s^2\hat{w}_0\big)_{z=h},
\end{alignat}
\end{subequations}
and the coefficient functions are given by:
\begin{subequations}\label{beta_mu}
\begin{alignat}{4}
\beta(s) &= \bar{G}s\big(\psi(sh) + \psi''(sh)\big) + k^2\bar{\gamma}_s s^2\psi'(sh),\quad 
\beta^{\diamond}(s) = \bar{G}s\big(\psi'(sh) + \psi'''(sh)\big) + k^2\bar{\gamma}_s s^2\psi''(sh),\\
\mu(s) &= \bar{G}s\big(3\psi'(sh) - \psi'''(sh)\big) + \bar{\gamma}_s s^2\psi(sh), \quad
\mu^{\diamond}(s) = \bar{G}s\big(3\psi''(sh) - \psi''''(sh)\big) + \bar{\gamma}_s s^2\psi'(sh).
\end{alignat}
\end{subequations}
These formulas define the solution of the boundary value problem \eqref{BVP1_Y}, from which we can calculate deformations as well as the stress and strain within the substrate, which are used in  the energy calculations of  \S\ref{s5p6}.  Note that in \eqref{beta_mu}, the shear modulus $G$ is constant in the case of the surface energy gradient outlined here ($\bar{G} = G$).  These expressions will be used throughout the remainder of the paper; in the case of a shear modulus gradient we will have $\bar{\gamma}_s = \gamma_s$.

\subsection{Stiffness Gradient}\label{s5p1}
Here we outline the static asymmetric two dimensional solution to the model for the case of a gradient in elastic shear modulus $G$ of the form:
\begin{equation}\label{Gvar}
G(x) = \bar{G} + a\tilde{G}(x) \qquad \tilde{G}(x) = \frac{2}{\pi}\arctan(x/L)
\end{equation}
where $\bar{G}$ is the average modulus, $a$ is a small parameter with $\bar{G}$ ($|a| \ll \bar{G}$), and $L$ is as defined previously. 
Following a similar procedure as \S\ref{s5p2}, we obtain the same boundary value problem \eqref{BVP0_Y} at $\mathcal{O}(1)$, and at $\mathcal{O}(a)$ we have:
 \begin{subequations}\label{BVP1}
\begin{alignat}{4}
 \nabla p_1 =& \ \bar{G}\Delta\langle u_1,w_1\rangle + \tilde{G}\Delta\langle u_0,w_0\rangle + \tilde{G}'\langle 2\partial_xu_0,\partial_xw_0+\partial_zu_0\rangle, \\
 \big[\bar{G}(\partial_zu_1 + \partial_xw_1) - k^2\gamma_s\partial_{xx}u_1\big]_{z=h} =& -\big[\tilde{G}(\partial_zu_0 + \partial_xw_0)\big]_{z=h}\label{BVP1_bc1}\\
\big[2\bar{G}\partial_zw_1 - p_1 - \gamma_s\partial_{xx}w_1\big]_{z=h} =& -\big[2\tilde{G}\partial_zw_0\big]_{z=h},\label{BVP1_bc2}
\end{alignat}
\end{subequations}
 Note that the first order correction boundary value problem \eqref{BVP1} is markedly different in structure from \eqref{BVP1_Y}.  By transforming   system \eqref{BVP1_Y}, we obtain  
\begin{equation}\label{BVP1_what}
(\partial_{zz} - s^2)^2\hat{w}_1 = \frac{is}{2\pi \bar{G}}\Big[-2\mathcal{F}\big[\tilde{G}'\big]*\big((\partial_{zz}-s^2)\hat{w}_0\big) + \Big(s\mathcal{F}\big[\tilde{G}'\big]\Big)*\big(s^{-1}(\partial_{zz} + s^2)\hat{w}_0\big)\Big],
\end{equation}
 together with $\hat{u}_1 = is^{-1}\partial_z\hat{w}_1$ from the  incompressibility condition \eqref{trace_strain}. 
 Solving, we obtain:
\begin{subequations}\label{uwhat1}
\begin{alignat}{2}
\hat{u}_1 &=\  iC_1(s)\psi'(sz) + iD_1(s)\psi''(sz) + i\hat{F}_1(s,z),\quad 
\hat{w}_1 = C_1(s)\psi(sz) + D_1(s)\psi'(sz) + \hat{F}_0(s,z)\label{what1},
\end{alignat}
\end{subequations}
where
$$
\hat{F}_k(s,z) = \frac{1}{2s^3}\int_0^z\Phi(s,z-\xi)\psi^{(k)}(s\xi)~d\xi, \quad k = 0,1,2,3
$$
and  $ \Phi(s,z) $ denotes the right hand side of \eqref{BVP1_what} and $\psi$ is as defined in \eqref{psidef}.  Switching the order of integration in $\hat{F}_k(s,z)$,   we obtain a useful version of the non-homogeneous solution to the first order correction terms:
\begin{equation}\label{Fhat}
\hat{F}_k(s,z) = \frac{i}{4\pi \bar{G}s^2}\int_{-\infty}^\infty \Big[-2\mathcal{F}\big[\tilde{G}'\big]\chi\big[(\partial_{zz}-\eta^2)\hat{w}_0(\eta,z),\psi^{(k)}\big] + \big(s\mathcal{F}\big[\tilde{G}'\big]\big)\chi\big[\eta^{-1}(\partial_{zz}+\eta^2)\hat{w}_0(\eta,z),\psi^{(k)}\big]\Big]~d\eta,
\end{equation}
 in which the operator $\chi$ is defined as
$$\chi\big[g(\eta,z),\psi^{(k)}(sz)\big] = \int_0^z g(\eta,z-\xi)\psi^{(k)}(s\xi)~d\xi.$$
Applying boundary conditions from \eqref{BVP0_Y}, \eqref{BVP1}, we obtain the same linear system \eqref{CDj_asym} as previously, but the right hand sides of the $\mathcal{O}(a)$ equations are now given by:
\begin{subequations}\label{MN1G}
\begin{alignat}{2}
M_1(s) &= \frac{i}{2\pi}\Big[\Big(\frac{\mathcal{F}\big[\tilde{G}'\big]}{s}\Big)*\big(s^{-1}(\partial_{zz}+s^2)\hat{w}_0\big)\Big]_{z=h} - \bar{G}s\big[\hat{F}_2(s,h) + \hat{F}_0(s,h)\big] - k^2\gamma_s s^2\hat{F}_1(s,h), \\
N_1(s) &= \frac{i}{2\pi}\Big[2\Big(\frac{\mathcal{F}\big[\tilde{G}'\big]}{s}\Big)*\partial_z\hat{w}_0 - s^{-1}\Big(\frac{\mathcal{F}\big[\tilde{G}'\big]}{s}\Big)*\big(s^{-1}(\partial_{zz}-s^2)\partial_z\hat{w}_0\big) + 2s^{-1}\mathcal{F}\big[\tilde{G}'\big]*\partial_z\hat{w}_0\Big]_{z=h} \nonumber \\
&\qquad - \bar{G}s\big[3\hat{F}_1(s,h) - \hat{F}_3(s,h)\big] - \gamma_s s^2\hat{F}_0(s,h),
\end{alignat}
\end{subequations}
while $M_0$ and $N_0$ are defined by \eqref{M0} and \eqref{N0} respectively.  Solving the linear system \eqref{CDj_asym} for $C_j$, $D_j$ ($j=0,1$) with right hand sides given by \eqref{MN1G} gives the transformed displacements $\hat{u}$ and $\hat{w}$ and thus the stress and strain within the solid substrate for the case of a gradient in shear modulus.  These calculations are used in \S\ref{s5p6}.  

\subsection{Elastic Energy and the Total Energy Functional}\label{s5p6}
Here we outline the procedure to calculate the total system energy as a function of contact angles $\theta_l$ and $\theta_r$ of a two dimensional droplet resting on an elastic substrate.  We then utilize MATLAB's optimization function {\texttt{fminsearch}} to determine the contact angles $\theta_l$ and $\theta_r$ that minimize this total energy for given system parameters and gradient applied to the substrate. 
 
 Elastic energy in the substrate is computed by the following integration over the solid domain $\Omega:$
$$\mathcal{E}_{elastic} = \frac{1}{2}\iint_\Omega \bar{\tau}\cdot\bar{\varepsilon}~d\Omega.$$
After manipulation we obtain a form suitable for the formulation on Fourier space:
\begin{equation}\label{Eelastic_FT}
\mathcal{E}_{elastic} = \frac{1}{\pi}\int_0^h \Big[\Big(\bar{G} + \frac{a}{2\pi}\big(\mathcal{F}\big[\tilde{G}\big]*\big)\Big)\big(\hat{\varepsilon}_{zz}*\hat{\varepsilon}_{zz} +\hat{\varepsilon}_{xz}*\hat{\varepsilon}_{xz}\big)\Big]_{s=0}~dz.
\end{equation}
\begin{figure}[h!]
\centering
\includegraphics[scale=.43]{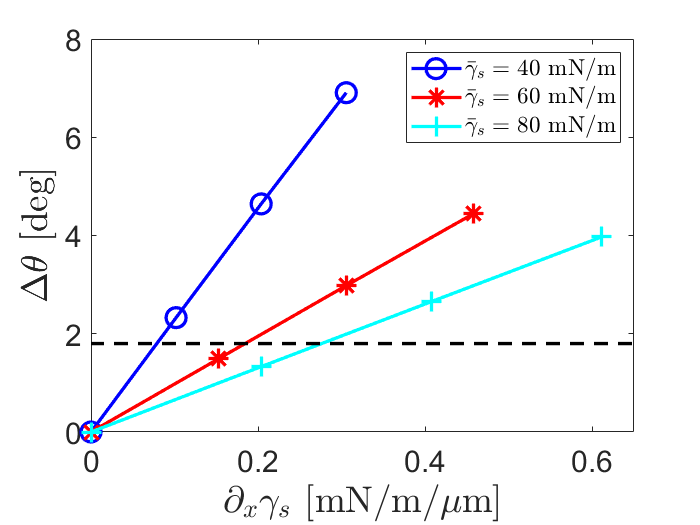}
\includegraphics[scale=.43]{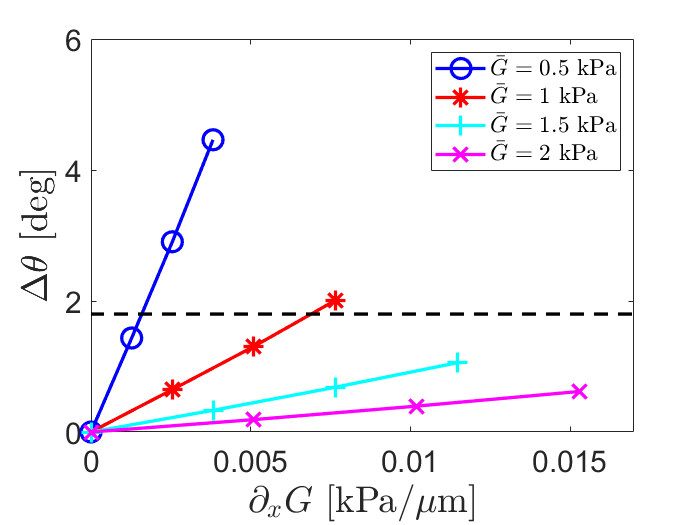}
\\
\includegraphics[scale=.43]{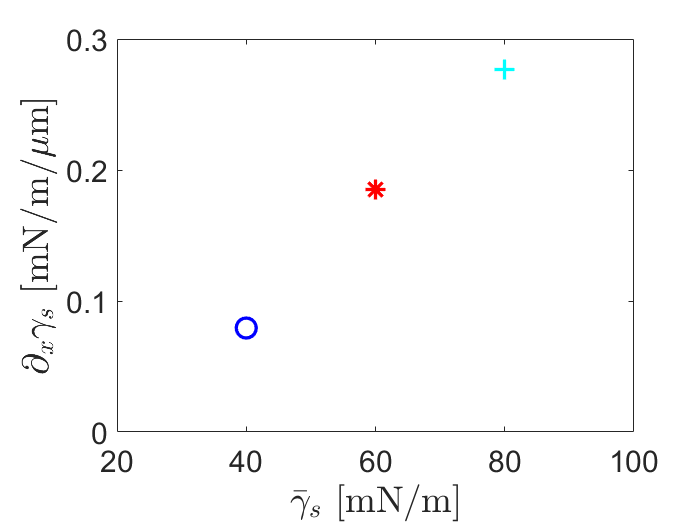}
\includegraphics[scale=.43]{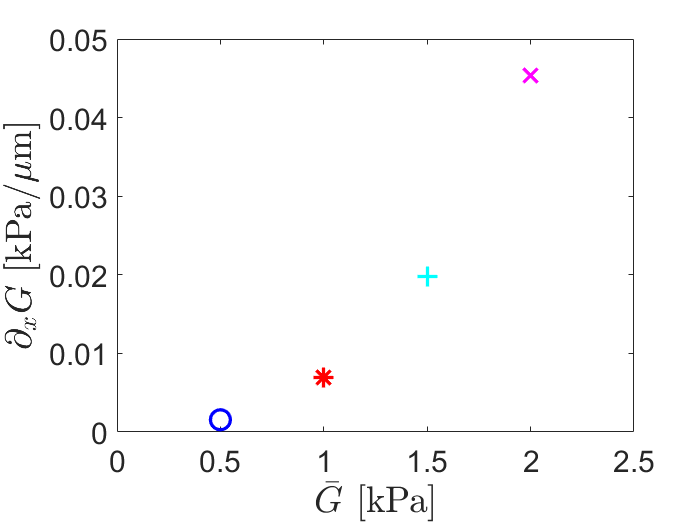}
\caption[Contact angle asymmetry as a function of gradient magnitude in both surface energy and shear modulus]{(a. \& b.) Contact angle asymmetry $\Delta\theta$ function of surface energy gradient (left) and shear modulus gradient (right).  Predicted threshold angle difference $\Delta\theta = 1.8^\circ$ plotted on dashed line.  Percent variation $|a|/\bar{(\cdot)}\times 100\%$ varied between 0 and 60\% in surface energy (left)/shear modulus (right).  (c. \& d.) Minimum surface energy gradient (left) and shear modulus gradient (right) needed to reach threshold angle difference $\Delta\theta = 1.8^\circ$.  Parameters: $h = 50~\mu$m, $A = 600\pi~\mu$m$^2$, $L = 50~\mu$m, $\gamma = 64$ mN/m, $G = 1$ kPa (left), $\gamma_s = 40$ mN/m (right).}
\label{fig:dalpha}
\end{figure}

The integrand of elastic energy \eqref{Eelastic_FT} is then numerically constructed via the outlined procedures in \S\ref{s5p2} and \S\ref{s5p1} where the deformation and strain transforms are approximated.  The total energy functional is then defined as
\begin{equation}\label{Edef}
\mathcal{E}_{total} = \mathcal{E}_{elastic} + \mathcal{E}_{surface},
\end{equation}
where the surface energy functional is given by
\begin{equation}\label{Esurf}
\mathcal{E}_{surface} = \gamma L_{drop} + \int_{-\infty}^\infty \gamma_s(x)\big[dl(x)-dx\big].
 \end{equation}
This formula relies on inverting the displacement transforms $\hat{u}$ and $\hat{w}$   to obtain the free surface profile. Then the solid surface energy, represented by the integral in \eqref{Esurf}, uses $dl(x),$ the differential arc length of the deformed free surface. The energy contribution of the liquid-gas interface is $\gamma L_{drop},$ where $L_{drop}$ is the arc length of the liquid-gas interface, calculated for prescribed contact angles $\theta_{l,r}$ and droplet area $A$, as calculated in \citet{r37}.  

The total energy minimization was done for typical ideal but physically realistic sets of parameters which could be experimentally tested.  Varying mean values of shear modulus $\bar{G}$ and surface energy $\bar{\gamma}_s$ for the stiffness gradient and surface energy gradient simulations respectively, we ran the minimization algorithm outlined in \S\ref{s5p6} and computed the contact angle difference $\Delta\theta = \theta_r - \theta_l$ as seen in Fig.~\ref{fig:dalpha}.  This is then compared to the benchmark contact angle difference from \citet{r5} of $\Delta\theta = 1.8^\circ$ necessary to drive droplet motion to obtain the necessary gradient for each mean value of modulus $\bar{G}$ or surface energy $\bar{\gamma}_s$.  We use these results to predict conditions for the induced spontaneous motion of the fluid droplet, for which we investigate dynamic velocity in \S\ref{dynamics}.

\section{Droplet Dynamics}\label{dynamics}
In this section we predict the dynamics of a droplet driven by a sufficiently large gradient, based on our results from \S\ref{asymmetric}. The velocity of the droplet is determined by calculating the rate at which total energy is released from the system and comparing this to the rate at which energy is dissipated from the substrate for a given velocity.  
 
In \S\ref{s6p1} we describe the procedure for obtaining the deformation field of the substrate caused by a two dimensional droplet migrating at velocity $v$ across the substrate surface.  In a short time $\Delta t$ the droplet moves a distance $\Delta x= v\Delta t.$ In this time, the viscoelastic solid has relaxed and the free surface has changed. With the droplet in the new position we can calculate  in \S\ref{s6p2} the new total energy and the solid energy dissipation  in the substrate. 
  Finally in  \S\ref{s6p4} we find the velocity $v$ by matching the rate of energy release to the rate of energy dissipation. The results are velocity predictions for both stiffness and surface energy gradients in the substrate. 
 
Throughout the section, we use the solid dissipation as an approximation for the total energy dissipation of the system, neglecting dissipation within the droplet.   This approximation, also employed  in \citet{r1}, is justified by experimental results by \cite{r39} in which motion of droplets over soft rubber is observed to be slower by several orders of magnitude compared to droplets on rigid surfaces, and it is shown that the kinetics are independent of liquid viscosity.  This phenomena, coined viscoelastic braking \citep{r39}, is  attributed to  
   the dominant  dissipation being from the solid rather than the liquid.  

\subsection{Moving Droplet Model}\label{s6p1}
We adopt the dynamic stress tensor given in \citet{r2}:
\begin{equation}\label{dyn_stress_tensor}
\tau_{ij}(x,z,t) = 2\int_{-\infty}^t \zeta(x,z,t-t')\dot{\varepsilon}_{ij}(x,z,t')~dt' - p(x,z,t)\delta_{ij},
\end{equation}
where $\zeta$ is the relaxation function of the soft solid.  For a reticulated polymer such as silicone gel, the response function has the power law form \citep{r2}:
\begin{equation}\label{relax}
\zeta(x,z,t) = G(x)\left(1+\frac{(t/t_v)^{-n}}{\Gamma(1-n)}\right),
\end{equation}
where $G(x)$ is the static shear modulus, $t_v$ is the viscous time scale, $n > 0$ is a fitting exponent and $\Gamma$ is the Gamma function.   
We introduce a temporal Fourier transform similar to the spatial transform defined previously:
$$\mathcal{F}_t\big[f(x,t)\big] = \int_{-\infty}^\infty f(x,t)e^{-i\omega t}dt = \breve{f}(x,\omega), \quad \mathcal{F}_t^{-1}\big[\breve{f}(x,\omega)\big] = \frac{1}{2\pi}\int_{-\infty}^\infty \breve{f}(x,\omega)e^{i\omega t}dt = f(x,t);$$
 further use of the spatial Fourier transform will be denoted by $\mathcal{F}_x\big[\cdot\big]$ or the $~\hat{ }~$ symbol.  For simplicity of notation we will write a twice transformed variable with a capital letter, for example:
$$\mathcal{F}_x\Big[\mathcal{F}_t\big[f(x,t)\big]\Big] = \mathcal{F}_t\Big[\mathcal{F}_x\big[f(x,t)\big]\Big] = F(s,\omega).$$
Applying the temporal Fourier transform to \eqref{dyn_stress_tensor} we obtain the temporally transformed stress tensor:
\begin{equation}\label{timetrans_tau}
\breve{\tau}_{ij} = 2g\breve{\varepsilon}_{ij} - \breve{p}~\delta_{ij},
\end{equation}
with complex shear modulus $g$ given by \citet{r2}:
\begin{equation}\label{complex_gmod}
g(x,\omega) = i\omega\int_0^\infty\zeta~e^{-i\omega t}dt = G(x)\big(1 + (i\omega t_v)^n\big).
\end{equation}
To adapt the solution to a moving droplet, we consider a droplet moving at a constant velocity $v > 0$.  The stress boundary conditions are rewritten in terms of the variable $x' = x - vt$ in the  moving reference frame:
\begin{subequations}\label{stressBC_moving}
\begin{alignat}{2}
\tau_{xz}(x',h) &= f_l\delta(x'+R) - f_r\delta(x' - R) + k^2\Upsilon\partial_{xx}u(x',h), \\
\tau_{zz}(x',h) &= \gamma\sin\theta_l\delta(x'+R) + \gamma\sin\theta_r\delta(x'-R) - \Pi H(R-|x'|) + \Upsilon\partial_{xx}w(x',h).
\end{alignat}
\end{subequations}
We consider quasi-static displacement in the substrate, as done in \citet{r2}:
$$\partial_j\tau_{ij} = 0, \quad \mbox{ so that}  \quad \mathcal{F}_t\big[\partial_j\tau_{ij}\big] = \partial_j\breve{\tau}_{ij} = 0,$$
with stress boundary conditions \eqref{stressBC_moving} at the free surface $z=h,$ along with fixed boundary conditions at the bottom surface $z=0$.  We proceed with only the zero order boundary value problem in the dynamic case, as the solution will be sufficient to predict energy and dissipation to leading order.  We define the dynamic boundary value problem formed by the fixed boundary conditions at $z=0$, stress boundary conditions given by \eqref{stressBC_moving} and manipulation of the quasi-static divergence-free stress tensor above, presented in temporally transformed space ($\bar{g}$ is the spatially averaged complex shear modulus, see \eqref{Gvar}, \eqref{complex_gmod}):
\begin{subequations}\label{movingBVP}
\begin{alignat}{4}
\nabla\breve{p} =& \bar{g}\Delta\langle \breve{u},\breve{w}\rangle,\\
 \left[\bar{g}(\partial_z\breve{u} + \partial_x\breve{w}) - k^2\bar{\gamma}_s\partial_{xx}\breve{u}\right]_{z=h} =& \mathcal{F}_t\left[f_l\delta(x'+R) - f_r\delta(x'-R)\right],\\
\left[2\bar{g}\partial_z\breve{w} - \breve{p} - \bar{\gamma}_s\partial_{xx}\breve{w}\right]_{z=h} =& \mathcal{F}_t\left[\gamma\big(\sin\theta_l\delta(x'+R) + \sin\theta_r\delta(x'-R)\big) - \Pi H(R-|x'|)\right].
\end{alignat}
\end{subequations}
We apply the spatial Fourier transform as in \S\ref{asymmetric} to solve \eqref{movingBVP}.  We use the following identity,
$$F(s,\omega) = \mathcal{F}_t\Big[\mathcal{F}_x\big[f(x-vt)\big]\Big] = 2\pi\hat{f}(s)\delta(\omega + sv),$$
where $\hat{f}(s)$ is the spatial Fourier transform of a function $f=f(x')$ in the moving reference frame.  Applying the spatial Fourier transform to \eqref{movingBVP} we obtain
$$\left(\partial_{zz} - s^2\right)^2W = 0, \qquad U = is^{-1}\partial_zW,$$
which when solved and applying the fixed boundary conditions gives us general solution
\begin{equation}\label{UW}
U(s,\omega,z) = iC(s,\omega)\psi'(sz) + iD(s,\omega)\psi''(sz), \quad
W(s,\omega,z) = C(s,\omega)\psi(sz) + D(s,\omega)\psi'(sz),
\end{equation}
where $\psi$ is again defined by \eqref{psidef}.  This leads to a linear system for Fourier coefficients $C$ and $D$:
\begin{align*}
C(s,\omega)\beta(s,\omega) + D(s,\omega)\beta^{\diamond}(s,\omega) =& 2\pi M_0(s)\delta(\omega+sv),\\
C(s,\omega)\mu(s,\omega) + D(s,\omega)\mu^{\diamond}(s,\omega) =& 2\pi N_0(s)\delta(\omega+sv),
\end{align*}
where formulas for $\beta,~\beta^{\diamond},~\mu,$ and $\mu^{\diamond}$ are as given in \eqref{beta_mu} but with the spatially averaged complex shear modulus $\bar{g}(\omega)$ in place of the average static shear modulus $\bar{G}$.
 
Next  we apply the inverse temporal Fourier Transform to obtain the spatial transforms of displacements $u$ and $w$ in the moving reference frame:
\begin{equation}\label{uwhat_moving}
\hat{w}(s,z) = \left[\frac{(\mu^{\diamond}M_0-\beta^{\diamond}N_0)\psi + (\beta N_0-\mu M_0)\psi'}{\beta\mu^{\diamond} - \beta^{\diamond}\mu}\right]_{\omega = -sv}, \qquad \hat{u}(s,z) = is^{-1}\partial_z\hat{w}(s,z).
\end{equation}
With these transform solutions \eqref{uwhat_moving}, we can proceed  in \S\ref{s6p2} to solve for the elastic energy necessary to calculate the rate at which energy is released from the system.  In addition we solve for the rate of solid energy dissipation; then comparison is used to predict droplet dynamics.

\subsection{Elastic Energy and Energy Dissipation Generated by the Moving Droplet}\label{s6p2}
The elastic energy for the dynamic model is calculated similarly to that derived in \S\ref{s5p6} using the dynamic stress tensor \eqref{dyn_stress_tensor}.  The dynamic elastic energy is given by
\begin{equation}\label{Eel_moving}
\mathcal{E}_{elastic}=\frac{\bar{G}}{\pi}\int_0^h\left[\hat{\varepsilon}_{zz}*\big((1+(-isvt_v)^n)\hat{\varepsilon}_{zz}\big) + \hat{\varepsilon}_{xz}*\big((1+(-isvt_v)^n\big)\hat{\varepsilon}_{xz}\big)\right]_{s=0}dz. 
\end{equation}
We use this formula to calculate the elastic energy contained in the substrate for a moving droplet traveling at constant speed $v$.  The integral in \eqref{Eel_moving} is calculated using Gaussian quadrature with the integrand  evaluated at Gaussian abscissae corresponding to the interval $z \in [0,h]$ and integrated numerically using the associated weights and integrand evaluations.
 
Note that equation \eqref{Eel_moving} generalizes the elastic energy for a given droplet velocity, and indeed is consistent with the zero order elastic energy formulation derived in \S\ref{asymmetric} for $v = 0$.   We construct the total energy functional $\mathcal{E}_{total}$ by including the new surface energy function from \eqref{Esurf},  and calculate the rate at which total energy is released from the system by taking a discrete derivative approximation of the total energy (in practice we use $\Delta t= 10^{-4} secs,$ but the results are insensitive to the choice of $\Delta t$):
\begin{equation}\label{Erelease}
\frac{d}{dt}\mathcal{E}_{total} \approx \frac{\mathcal{E}_{total}(\Delta t) - \mathcal{E}_{total}(0)}{\Delta t}.
\end{equation}
Similar to the elastic energy in the system, we seek to compute the rate at which energy is dissipated $\mathcal{P}_{solid}$ within the substrate.  Balancing this with the rate at which energy is released from the system due to migration will give us the predicted velocity $v$ of the droplet for a given gradient in substrate properties.  The formula for solid dissipation is given by \citet{r1,r32}:
 $$\mathcal{P}_{solid} = \frac{1}{2}\iint_{\Omega}\bar{\tau}\cdot\dot{\bar{\varepsilon}}~d\Omega,$$
 where we manipulate similar to the definition of elastic energy to obtain:
\begin{equation}\label{Psolid2}
\mathcal{P}_{solid}=\frac{\bar{G}v}{\pi}\int_0^h\left[\big(-is\hat{\varepsilon}_{zz}\big)*\big((1+(-isvt_v)^n)\hat{\varepsilon}_{zz}\big) + \big(-is\hat{\varepsilon}_{xz}\big)*\big((1+(-isvt_v)^n)\hat{\varepsilon}_{xz}\big)\right]_{s=0}dz.
\end{equation}
With traveling displacement transforms \eqref{uwhat_moving} defined in \S\ref{s6p1}, we can numerically construct and evaluate the integrand of \eqref{Psolid2} at Gaussian nodes on interval $z \in [0,h]$ and evaluate \eqref{Psolid2} using Gaussian quadrature as done for \eqref{Eel_moving}.  With these computations we can solve for the velocity $v$ such that the rate at which energy is released by migration \eqref{Erelease} matches that of dissipation within the solid \eqref{Psolid2}.  Furthermore we can assign gradients in surface energy or shear modulus above the critical gradient necessary to induce motion to predict the droplet velocity for larger gradients.

\subsection{Dynamics Results}\label{s6p4}
\begin{figure}[h!]
\centering
\includegraphics[scale=.43]{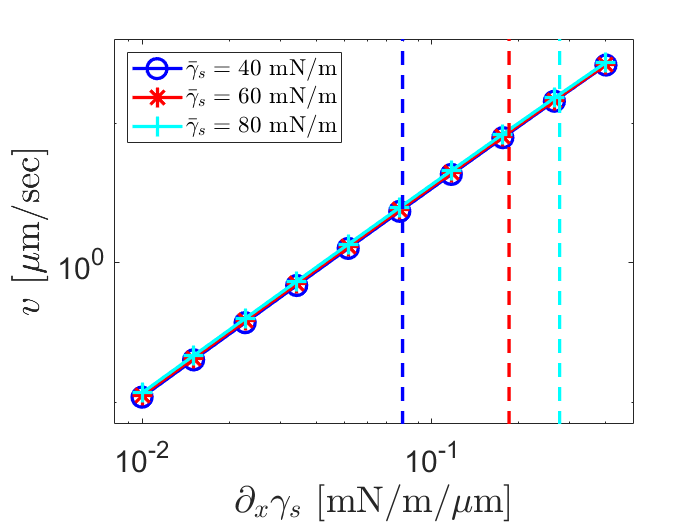}
\includegraphics[scale=.43]{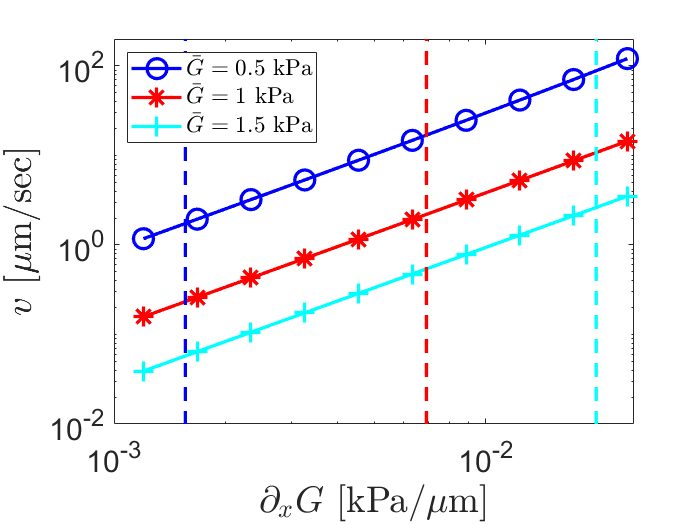}
\caption{Dynamic velocity predictions of fluid droplets resulting from gradients in surface energy $\gamma_s$ (left) and shear modulus $G$ (right).   
Vertical dashed lines show the minimum gradient necessary to induce motion for the given average surface energy $\bar{\gamma}_s$ (left) or shear modulus $\bar{G}$ (right).  Parameters: $h = 50~\mu$m, $A = 600\pi~\mu$m$^2$, $\gamma = 64$ mN/m, $L = 50~\mu$m, $G = 1$ kPa (left), $\gamma_s = 40$ mN/m (right), $t_{v} = 0.03$ sec, $n = 2/3$.}
\label{fig:velplots}
\end{figure}

With the formulation of energy release rate \eqref{Erelease} and energy dissipation \eqref{Psolid2} from \S\ref{s6p2}, we can now predict the velocity $v$ of a droplet for which motion has been induced by equating the rate of energy release with the rate at which energy is dissipated to heat, for which we solve the equation
\begin{equation}\label{EPcomp}
\frac{d}{dt}\mathcal{E}_{total}(v;\tilde{G},\tilde{\gamma}_s) = \mathcal{P}_{solid}(v;\tilde{G},\tilde{\gamma}_s).
\end{equation}
The solution $v$ of \eqref{EPcomp} is obtained by a bisection algorithm, and
solution curves are shown in Fig.~\ref{fig:velplots}.  These plots indicate that the predicted velocities in the case of the surface energy gradient are largely universal as a function of the mean surface energy $\bar{\gamma}_s$ while the velocities in the case of the shear modulus gradient exhibit distinct behavior for different mean shear moduli $\bar{G}$.  A power law fit of the form $v \propto \big(\partial_x(\cdot)\big)^b$ gives exponents $b_{\gamma_s} \approx 1.48$ and $b_{G} \approx 1.51$ for the surface energy and shear modulus gradients respectively.

\section{Discussion}\label{Discussion}
In Fig.~\ref{fig:dalpha}, we observe that in the parameter ranges tested, the shear modulus gradient is less effective than the surface energy gradient in driving the contact angle asymmetry to the threshold level for inducing droplet motion.  
In the figure we can see that only average shear moduli of $\bar{G} = 1$ kPa or lower are successful in generating a contact angle difference sufficient to drive droplet motion according the the benchmark threshold of 1.8$^\circ$ (simulations calculated up to $|a|/\bar{G} = 0.6$, $L = 50~\mu$m).

We can benchmark these results against experimentally-accessible soft substrates.
By inter-diffusing soft gel networks with differing moduli,  \citet{r27} are  able to produce a total variation in modulus of approximately $90\%$ over roughly 10 cm with average modulus $\bar{G} \approx 35$~kPa.
A solid with this average shear modulus and modulus gradient, as indicated by our results, 
would be insufficient to drive droplet motion. 
Similarly, gradients in ultra-violet intensity during the curing of PDMS gels, as discussed in \cite{r25}, do not create a modulus gradient large enough to drive motion of a fluid droplet.
The most promising option, by \cite{r55}, is to cure  styrenated gelatin using visible light. This results in materials with stiffness $\bar{G} \approx 1$~kPa and a gradient around 0.04 kPa$/\mu$m, close to our predicted limit of motion.
This suggests that advances must be made to create substrates which exhibit both extremely soft elastic moduli as well as sharp moduli gradients to make motion induced by a substrate gradient attainable.  For these reasons, we believe that the stiffness gradient is likely a physically 
unrealistic method for inducing spontaneous droplet motion.  
 
Most previous durotaxis experiments have been done with either a substrate stiffer than the threshold of our calculation \citep{r57} or with the stiffness gradient smaller than the threshold of our calculation \citep{r56}. In addition, cell durotaxis \citep{r4} and droplet motion on thickness gradients \citep{r5} exhibit dynamics in opposite directions from each other. Therefore, we believe that the elastocapillarity plays an insignificant role in cellular durotaxis. Instead, cells sense the substrate stiffness and change the cytoskeleton, which induces the cell motion.

On the contrary, the capability to generate a contact angle asymmetry using a surface energy gradient is experimentally feasible.  In \citet{r41}, by exposing the surface to the diffusing front of a vapor of decyltrichlorosilane (Cl$_3$Si(CH$_2$)$_9$CH$_3$), a gradient in solid surface energy was generated causing a contact angle asymmetry of 6-8$^\circ$.  This contact angle asymmetry was sufficient to cause droplets of water to migrate uphill on a rigid surface.  
Supplying asymmetry of the magnitude capable by methods in \citet{r41} is sufficient to meet the benchmark asymmetry threshold of 1.8$^\circ$ \citep{r5}.    
 
In Fig.~\ref{fig:velplots} we see the predicted droplet velocities resulting from gradients in surface energy $\gamma_s$ or shear modulus $G$.  Here we observe that droplet velocities increase as the shear modulus decreases, while velocities as the result of surface energy gradients are universal given the gradient is sufficient to spontaneously induce droplet motion.  This indicates that in the case of the surface energy gradient, the mean stiffness determines the gradient necessary to induce motion but is not a factor in the resulting dynamics.  On the other hand, dynamics as a result of a shear modulus gradient is largely influenced by the mean shear modulus.  We believe the reason for the different behaviors is the result of the energy minimizing conditions for each substrate property.  By adjusting the shear modulus, the deformation of the substrate and therefore the elastic and surface energies of the system are drastically altered.  However, the deformation and elastic energy of the substrate are highly independent of the surface energy.  While the total surface energy as a whole is largely affected by adjusting the surface energy $\gamma_s$, the total {\em change} in surface energy from the undeformed to deformed state will also be independent of the mean surface energy.
 
Our results suggest that spontaneous droplet motion as a result of an elastic modulus gradient is currently infeasible given current experimental capabilities.  However, we see from the results of Figs.~\ref{fig:dalpha} and \ref{fig:velplots} that dynamics as a result of gradients in elastic modulus drastically benefit from extremely low moduli in both the gradient required to induce motion and the resulting droplet velocity.  For the case of surface energy gradients, our results indicate that droplet motion over soft solids as a result of a surface energy gradient is currently feasible despite larger magnitude energy dissipation from the soft solid.  The dynamics of droplets exposed to a surface with a surface energy gradient benefit from lower surface energy values, which reduces the gradient necessary to induce motion. 
 
\section*{Acknowledgment}

This work was supported by National Science Foundation grants DMS-151729 and DMR-1608097.


\begin{thebibliography}{}

\bibitem[Ahmed(2014)]{r10}
{\sc Ahmed, G., Mathieu, S., Jermy, M., Taylor, M.} (2014) Modeling the Effects of Contact Angle Hysteresis on the Sliding of Droplets Down Inclined Surfaces. {\em European Journal of Mechanics B/Fluids}, \textbf{48}, 218-230.

\bibitem[Andreotti(2016)]{r40}
{\sc Andreotti, B., B\"aumchen, O., Boulogne, F., Daniels, K.~E., Dufresne, E.~R., Perrin, H., Salez, T., Snoeijer, J.~H., Style, R.~W.} (2016) Soft Capillarity: When and How Does Surface Tension Deform Soft Solids? {\em Soft Matter}, \textbf{12} 2993-2996.

\bibitem[Andreotti(2016)]{r41}
{\sc Andreotti, B., Snoeijer, J.~H.} (2016) Soft Wetting and the Shuttleworth Effect, at the Crossroads Between Thermodynamics and Mechanics. {\em Europhysics Letters}, \textbf{113}, 66001.

\bibitem[Bardall(2018)]{r24}
{\sc Bardall, A., Daniels, K.~E., Shearer, M.} (2018) Deformation of an Elastic Substrate Due to a Resting Sessile Droplet. {\em European Journal of Applied Mathematics}, \textbf{29 (2)}, 281-300.

\bibitem[Bico(2018)]{r30}
{\sc Bico, J., Reyssat, \'E., Roman, B.} (2018) Elastocapillarity: When Surface Tension Deforms Elastic Solids, {\em Annual Reviews}, \textbf{50}, 629-659.

\bibitem[Bostwick(2014)]{r52}
{\sc Bostwick, J.~B., Shearer, M., Daniels, K.~E.} (2014) Elastocapillary Deformations on Partially-Wetting Substrates: Rival Contact-Line Models. {\em Soft Matter}, \textbf{10}, 7361-7369.

\bibitem[Bueno(2017)]{r53}
{\sc Bueno, J., Bazilevs, Y., Juanes, R., Gomez, H.} (2017) Droplet Motion Driven by tensotaxis. {\em Extreme Mechanics Letters}, \textbf{13}, 10-16.

\bibitem[Bueno(2018)]{r4}
{\sc Bueno, J., Bazilevs, Y., Juanes, R., Gomez, H.} (2018) Wettability Control of Droplet Durotaxis. \em{Soft Matter}, \textbf{14}, 1417-1426.

\bibitem[Chaudhury(1992)]{r41}
{\sc Chaudhury, M.~K., Whitesides, G.~M.} (1992) How to Make Water Run Uphill. {\em Science}, \textbf{256}, 1539-1541.

\bibitem[Crowe-Willoughby(2010)]{r27}
{\sc Crowe-Willoughby, J.~A., Weiger, K.~L., \"Ozcam, A.~E., Genzer, J.} (2010) Formation of Silicone Elastomer Networks Films with Gradients in Modulus. \textit{Polymer}, \textbf{51 (3)}, 763-773.

\bibitem[Dervaux(2015)]{r19}
{\sc Dervaux, J., Limat, L.} (2015) Contact Lines on Soft Solids with Uniform Surface Tension: Analytical Solutions and Double Transition for Increasing Deformability. {\em Proceedings of the Royal Society A}, \textbf{471}, 2176.

\bibitem[Dhir(2004)]{r42}
{\sc Dhir, V., Gao, D., Morley, N.~B.} (2004) Understanding Magnetic Field Gradient Effect from a Liquid Metal Droplet Movement. {\em Journal of Fluids Engineering}, \textbf{126}, 120-124.

\bibitem[Herde(2013)]{r37}
{\sc Herde, D.} (2013) {\em Contact Line Dynamics on Heterogeneous Substrates.}  (Doctoral dissertation), \textit{Georg-August University School of Science}.

\bibitem[Hourlier-Fargette(2017)]{r9}
{\sc Hourlier-Fargette, A., Anthowiak, A., Chateauminois, A., Neukirch, S.} (2017) Role of Uncrosslinked Chains in Droplets Dynamics on Silicone Elastomers. {\em Soft Matter}, \textbf{13}, 3484-3491.

\bibitem[Hourlier-Fargette(2018)]{r34}
{\sc Hourlier-Fargette, A., Dervaux, J., Antkowiak, A., Neukirch, S.} (2018) Extraction of Silicone Uncrosslinked Chains at Air-Water-Poydimethylsiloxane Triple Lines. {\em Langmuir}, \textbf{34 (41)}, 12244-12250.

\bibitem[Hui(2014)]{r44}
{\sc Hui, C.~Y., Jagota, A.} (2014) Deformation Near a Liquid Contact Line on an Elastic Substrate.  {\em Proceedings of the Royal Society A}, \textbf{470}, 20140085.

\bibitem[Jerison(2011)]{r17}
{\sc Jerison, E.~R., Xu, Y., Wilen, L.~A., Dufresne, E.~R.} (2011) Deformation of an Elastic Substrate by a Three-Phase Contact Line. {\em Physical Review Letters}, \textbf{106}, 186103.

\bibitem[Karpitschka(2015)]{r2}
{\sc Karpitschka, S., Das, S., van Gorcum, M., Perrin, H., Andreotti, B., Snoeijer, J.~H.} (2015) Droplets Move Over Viscoelastic Substrates by Surfing a Ridge. {\em Nature Communications}, \textbf{6}, 7891.

\bibitem[Kourasi(2018)]{r20}
{\sc Koursari, N., Ahmed, G., Starov, V.~M.} (2018) Equilibrium Droplets on Deformable Substrates: Equilibrium Conditions. {\em Langmuir}, \textbf{34 (19)}, 5672-5677.

\bibitem[Kidoaki(2008)]{r57}
{\sc Kidoaki, S., Matsuda, T.} (2008) Microelastic gradient gelatinous gels to induce cellular mechanotaxis. {\em Journal of Biotechnology}, \textbf{133 (2)}, 225-230.

\bibitem[Limat(2012)]{r45}
{\sc Limat, L.} (2012) Straight Contact Lines on a Soft, Incompressible Solid.  {\em European Phys. Journal E.}, \textbf{35}, 1-13.

\bibitem[Long(1996)]{r1}
{\sc Long, D., Ajdari, A., Leibler, L.} (1996) Static and Dynamic Wetting Properties of Thin Rubber Films. {\em Langmuir}, \textbf{12 (21)}, 5221-5230.

\bibitem[Lubbers(2014)]{r11}
{\sc Lubbers, L.,~A., Weijs, J.~H., Botto, L., Das, S.} (2014) Drops on Soft Solids: Free Energy and Double Transition of Contact Angles. {\em Journal of Fluid Mechanics}, \textbf{747}

\bibitem[Moriyama(2019)]{r55}
{\sc Moriyama, K., Kidoaki, S.} (2019) Cellular Durotaxis Revisited: Initial-Position-Dependent Determination of the Threshold Stiffness Gradient to Induce Durotaxis. {\em Langmuir}, \textbf{35 (23)}, 7478-7486.

\bibitem[Onuki(2005)]{r46}
{\sc Onuki, A., Kanatani, K.} (2005) Droplet Motion with Phase Change in a Temperature Gradient. {\em Physical Review E}, \textbf{72}, 27844.

\bibitem[Palchkesko(2012)]{r26}
{\sc Palchesko, R.~N., Zhang, L., Sun, Y., Feinber, A.~W.} (2012) Development of Polydimethylsiloxane Substrates with Tunable Elastic Modulus to Study Cell Mechanobiology in Muscle and Nerve. {\em PLoS ONE}, \textbf{7 (12)}, e51499.

\bibitem[Park(2014)]{r16}
{\sc Park, S.~J., Weon, B.~M., Lee, J.~S., Kim, J., Je, J.~H.} (2014) Visualization of Asymmetric Wetting Ridges on Soft Solids with X-ray Microscopy. {\em Nature Communications}, \textbf{5}, 4369.

\bibitem[Park(2017)]{r7}
{\sc Park, S.~J., Bostwick, J.~B., De Andrade, V., Je, J.~H.} (2017) Self-spreading of the Wetting Ridge During Stick-slip on a Viscoelastic Surface. {\em Soft Matter}, \textbf{13}, 8331-8336.

\bibitem[Schulman(2018)]{r23}
{\sc Schulman, R.~D., Trejo, M., Salez, T., Rapha\"el, E., Dalnoki-Veress, K.} (2018) Surface Energy of Strained Amorphous Solids. {\em Nature Communications}, \textbf{9}, 982.

\bibitem[Shanahan(1995)]{r39}
{\sc Shanahan, M.~E.~R., Carr\'e, A.} (1995) Viscoelastic Dissipation in Wetting and Adhesion Phenomena.  {\em Langmuir}, \textbf{11 (4)}, 1396-1402.

 \bibitem[Snoeijer(2013)]{r31}
{\sc Snoeijer, J.~H., Andreotti, B.} (2013) Moving Contact Lines: Scales, Regimes, and Dynamical Transitions. {\em Annual Reviews,} \textbf{45}, 269-292.

\bibitem[Snoeijer(2018)]{r12}
{\sc Snoeijer, J.~H., Rolley, E., Andreotti, B.} (2018) Paradox of Contact Angle Selection on Stretched Soft Solids. {\em Physical Review Letters}, \textbf{121}, 068003.

\bibitem[Soutas-Little(1999)]{r38}
{\sc Soutas-Little, R.~W.} (1999)
\newblock {\em Elasticity.}
Dover Publications.

\bibitem[Stricher(2016)]{r25}
{\sc Stricher, A., Rinaldi, R.~G., Machado, G., Chagnon, G., Favier, D., Chazeau, L., Ganachaud, F.} (2016) Light-induced Bulk Architecturation of PDMS Membranes. {\em Macromolecula Materials and Engineering}, \textbf{301 (10)}, 1151-1157.

\bibitem[Style(2012)]{r15} 
{\sc Style, R.~W., Dufresne, E.~R.} (2012) Static Wetting on Deformable Substrates, From Liquids to Soft Solids. {\em Soft Matter}, \textbf{8}, 7177-7184.

\bibitem[Style(2013a)]{r54}
{\sc Style, R.~W., Boltyanskiy, R., Che, Y., Wettlaufer, J.~S., Wilen, L.~A., Dufresne, E.~R.} (2013a) Universal Deformation of Soft Substrates Near a Contact Line and the Direct Measurement of Solid Surface Stresses. {\em Physical Review Letters}, \textbf{110}, 066103.

\bibitem[Style(2013b)]{r5}
{\sc Style, R.~W., Che, Y., Park, S.~J., Weon, B.~M., Je, J.~H.,  Hyland, C., German, G.~K., Power, M.~P., Wilen, L.~A., Wettlaufer, J.~S., Dufresne, E.~R.} (2013b) Patterning Droplets with Durotaxis. {\em Proceedings of the National Academy of Sciences}, \textbf{110 (31)}, 12541-12544.

\bibitem[Style(2013c)]{r18}
{\sc Style, R.~W., Hyland, C., Boltyanskiy, R., Wettlaufer, J.~S., Dufresne, E.~R.} (2013c) Surface Tension and Contact with Soft Elastic Solids. {\em Nature Communications}, \textbf{4}, 2728.

\bibitem[Style(2017)]{r29}
{\sc Style, R.~W., Jagota, A., Hui, C., Dufresne, E.R.} (2017) Elastocapillarity: Surface Tension and the Mechanics of Soft Solids. {\em Annual Reviews}, \textbf{8}, 99-118.

\bibitem[Style(2018)]{r21}
{\sc Style, R.~W., Xu, Q.} (2018) The Mechanical Equilibrium of Soft Solids with Surface Elasticity. {\em Soft Matter}, \textbf{14}, 4569-4576.

\bibitem[Sun(2019)]{r22}
    {\sc   Sun, Q.,   Wang, D.,   Li, Y.,    Zhang, J.,    Ye, S.,    Cui, J.,    Chen, L.,    Wang, Z.,   Butt, H.-J.,   Vollmer, D.,  Xu  , X.} (2019) Surface charge printing for programmed droplet transport.  {\em Nature Materials}, DOI: 10.1038.s41563-019-0440-2.

 \bibitem[Theodorakis(2017)]{r3}
{\sc Theodorakis, P.E., Egorov, S.~A., Milchev, A.} (2017) Stiffness-guided Motion of a Droplet on a Solid Substrate. {\em Journal of Chemical Physics}, \textbf{146}, 244705.

\bibitem[Tschoegl(2002)]{r32}
{\sc Tschoegl, N.~W.} (2002)
\newblock {\em The Phenomenological Theory of Linear Viscoelastic Behavior},
Springer, Berlin, Heidelberg.

\bibitem[van \, Gorcum(2019)]{r28} {\sc van Gorcum, M., Karpitschka,  S., Andreotti,  B., and Snoeijer, J. H.} (2019)
Spreading on viscoelastic solids: Are contact angles selected by Neumann's law? {\em arXiv: 1907.08067v1.}

\bibitem[Vou\'e(2003)]{r8} 
{\sc Vou\'e, M., Rioboo, R., Bauthier, C., Conti, J., Carlot, M., De Coninck, J.} (2003) Dissipation and Moving Contact Lines on Non-rigid Substrates. {\em Journal of the European Ceramic Society}, \textbf{23 (15)}, 2769-2775.

\bibitem[Wong(2003)]{r56} 
{\sc Wong, J.~Y., Velasco, A., Rajagopalan, P., Pham, Q.} (2003) Directed Movement of Vascular Smooth Muscle Cells on Gradient-Compliant Hydrogels. {\em Langmuir}, \textbf{19 (5)}, 1908-1913.

\bibitem[Xu(2018)]{r22}
{\sc Xu, Q., Style, R.~W., Dufresne, E.~R.} (2018) Surface Elastic Constants of a Soft Solid. {\em Soft Matter}, \textbf{14}, 916-920.

\bibitem[Zhao(2018)]{r6}
{\sc Zhao, M., Dervaux, J., Narita, T., Lequeux, F., Limat, L., Roch\'e, M.} (2018) Geometrical Control of Dissipation During the Spreading of Liquids on Soft Solids. {\em Proceedings of the National Academy of Sciences}, \textbf{115 (8)}, 1748-1753.

    
\end{thebibliography}
\end{document}